\documentclass[aps,prb,reprint,preprintnumbers,superscriptaddress,amsmath,amssymb,bibnotes,longbibliography]{revtex4-2}
\usepackage{graphicx}
\usepackage{dcolumn}
\usepackage{epstopdf}
\usepackage{bm}
\usepackage{flushend}
\usepackage[pdfstartview=FitH, CJKbookmarks=true, bookmarksnumbered=true, bookmarksopen=true, colorlinks, pdfborder=001, linkcolor=blue, anchorcolor=blue, citecolor=blue,urlcolor=blue,breaklinks]{hyperref}
\usepackage[mathlines]{lineno}
\usepackage {gensymb} 
\usepackage{amssymb}
\makeatletter
\newcommand{\Rmnum}[1]{\expandafter\@slowromancap\romannumeral #1@}
\makeatother
\usepackage{booktabs}
\usepackage{float}
\usepackage{graphicx}
\usepackage{color}
\begin{document}
\title{Inelastic neutron scattering and muon spin relaxation investigations of the deuterated Kondo lattices CeNiSnD$ _x $ }
\author{X. Y. Zheng}
\affiliation{Center for Correlated Matter and School of Physics, Zhejiang University, Hangzhou 310058, China}
\author{D. T. Adroja}
\email[Corresponding author: ]{devashibhai.adroja@stfc.ac.uk}
\affiliation{ISIS Facility, STFC Rutherford Appleton Laboratory, Harwell Ocford, Oxfordshire OX11 0QX, United Kingdom}
\affiliation  {Highly Correlated Matter Research Group, Physics Department, University of Johannesburg, P.O. Box 524, Auckland Park 2006, South Africa}
\author{B. Chevalier}
\affiliation{ICMCB, CNRS (UPR 9048), Avenue du Dr. A. Schweitzer, 33608 Pessac, France}
\author{Z. Y. Shan}
\affiliation{Center for Correlated Matter and School of Physics, Zhejiang University, Hangzhou 310058, China}
\author{A. D. Hillier}
\affiliation{ISIS Facility, STFC Rutherford Appleton Laboratory, Harwell Ocford, Oxfordshire OX11 0QX, United Kingdom}
\author{H. Q. Yuan}
\affiliation  {Center for Correlated Matter and School of Physics, Zhejiang University, Hangzhou 310058, China}
\affiliation{Collaborative Innovation Center of Advanced Microstructures, Nanjing 210093, China}
\affiliation  {State Key Laboratory of Silicon Materials, Zhejiang University, Hangzhou 310058, China}
\author{M. Smidman}
\email[Corresponding author: ]{msmidman@zju.edu.cn}
\affiliation{Center for Correlated Matter and School of Physics, Zhejiang University, Hangzhou 310058, China}
\date{\today}

\begin{abstract}
CeNiSn is a Kondo semimetal where a gap opens at low temperatures due to hybridization between 4$f$ and conduction electrons, but a full insulating state fails to develop. Upon the insertion of hydrogen, long range magnetic order is induced. Here we report zero-field muon-spin relaxation and inelastic neutron scattering measurements of polycrystalline samples of the deuterides CeNiSnD$_x$ ($x$=1.0, 1.8). The muon-spin relaxation results confirm magnetic ordering in the whole sample of CeNiSnD below around 4.7~K, while  inelastic neutron scattering reveals  two well-defined crystalline-electric field (CEF) excitations at around 13~meV and 34~meV in CeNiSnD, and 5~meV and 27~meV for CeNiSnD$_{1.8}$. These  results suggest that hydrogenation  leads to the localization of the Ce-4$f$ electrons, giving rise to long-range magnetic order. We propose CEF level schemes for both systems, which predict a ground state moment of 0.96$\mu_{\rm B}$/Ce within the $ab$-plane for CeNiSnD$_{1.8}$ and a saturated moment of 1.26$\mu_{\rm B}$/Ce along the easy $c$ axis for CeNiSnD, that account for the observed magnetic properties.
\begin{description}
\item[PACS number(s)]

\end{description}
\end{abstract}

\maketitle
\section{INTRODUCTION}
 Ce-based Kondo lattices can exhibit various electronic phases arising from competition between the Ruderman-Kittel-Kasuya-Yosida (RKKY) interaction and the Kondo effect, including complex magnetic order, unconventional superconductivity, strange metal behavior and quantum criticality \cite{HFQS,HFSC,Weng2016}. The relative strengths of these interactions can be adjusted by non-thermal parameters such as hydrostatic pressure, magnetic fields, and chemical doping, which can often readily tune the ground states of Kondo lattice systems. Hydrogenation is one such means of tuning Kondo lattices, whereby the insertion of hydrogen generally expands the lattice, corresponding to a negative chemical pressure which decreases the hybridization strength. This can either induce magnetic ordering  in otherwise non-magnetic heavy fermions \cite{CeTiGeH2019,CeIrSbH2009,CeRuSiH2008} or tune the ordering temperatures of Kondo magnets \cite{CePdSnHandCePdInH,CeScSiH}. On the other hand, strong bonding between the rare-earth and hydrogen ions can modify the electronic structure, resulting in the delocalization of Ce-4$f$ electrons, and change the ground state from magnetic ordering to one with spin fluctuations or intermediate valence
\cite{CeCoSiH,CeCoSiHandCeCoGeH,CeCoGeH,GdScGeH}.

CeNiSn is an unusual heavy fermion system, whereby transport measurements evidence the opening of a gap due to Kondo hybridization between the 4$f$ and conduction electrons, but a full Kondo insulating state fails to develop \cite{TkCeNiSn,PhysRevB.45.5740,notfullyinsulating,Izawa1999}, which had been ascribed to a V-shaped density of states at the Fermi level arising from highly anisotropic or even nodal Kondo hybridization \cite{Tunnelingforvshape,NMRcenisn,Nakamura1994,Nersteffect,FLtheory,Moreno2000}. Furthermore, no magnetic ordering is detected down to 0.1~K also evidencing a strong Kondo effect \cite{KYOGAKU1991}. Recently,  CeNiSn was also proposed to be a topological Kondo semimetal with M{\"o}bius-twisted surface states \cite{mobius,PhysRevB.99.235105}, while the high temperature band structure may also correspond to a Dirac nodal-loop semimetal with hourglass-type bulk band crossings \cite{Nam2019,ARPES}.  

Meanwhile, inelastic neutron scattering (INS) at low temperatures reveals unusual magnetic excitations together with a spin-gap that appear to exist only at certain momentum transfers \textbf{Q} \cite{T.E.MasonINS,HiroakiKadowaki1994,T.J.SatoINS}, and it is not settled whether these primarily correspond to antiferromagnetic correlations \cite{HiroakiKadowaki1994,T.J.SatoINS}, interband transitions from the heavy renormalized Fermi surface with a partial hybridization gap \cite{FLtheory,INShighfield}, or alternatively a heavy-fermion spin liquid arising due to hybridization between crystalline-electric field (CEF) excitations and conduction electrons \cite{Kagan1993,PhysRevB.55.12348}. Moreover, an additional magnetic excitation was detected in CeNiSn at higher energy transfers of around 40~meV \cite{CEFCeNiSn,possibleCEFCeNiSn}, which may correspond to a CEF excitation broadened by the Kondo hybridization, while doping with Cu, Pd and Pt for Ni leads to the appearance of two well-defined CEF excitations \cite{INSofdopedCeNiSn,CePdSn,CEFCePtSnNidoped}.
 
Two hydrides derived from CeNiSn have been reported, CeNiSnH and CeNiSnH$_{1.8}$, where CeNiSnH has the same orthorhombic TiNiSi-type structure as CeNiSn and orders antiferromagnetically below 4.5-5.1 K  \cite{CHEVALIER2004576,CeNiSnH,PNDofCeNiSnD}, while CeNiSnH$_{1.8}$  has a hexagonal ZrBeSi-type structure and has a ferromagnetic transition at $T_{\rm C}$ = 7 K \cite{CHEVALIER2004576,CeNiSnH1.8}. In both cases, hydrogenation induces a small expansion of the unit cell volume, leading to  magnetic ordering, while the influence  of the Kondo interaction can be inferred by the reduced magnetic entropy at the ordering temperature. This influence is more prominent in CeNiSnH$_{1.8}$ than CeNiSnH, since the former also exhibits a Kondo-like increase of the resistivity with decreasing temperature, and has a larger Sommerfeld coefficient $\gamma_{\rm ele}$ \cite{CHEVALIER2004576}. The ordered magnetic moment of CeNiSnH is estimated to be 1.37$\mu_{\rm B}$/Ce along the $c$ axis from neutron diffraction \cite{PNDofCeNiSnD}. The relatively small ordered moments in rare-earth compounds may be due to the reduced moments of the ground state Kramers doublet arising from the splitting of the ground state $J = 5/2$ multiplet, while Kondo hybridization can further reduce the values. Disentangling these effects requires probing the CEF excitations and determining the CEF level schemes.

 It is therefore of particular interest to characterize the magnetic properties of CeNiSnH and CeNiSnH$_{1.8}$ using microscopic techniques, and in particular to understand the evolution of the magnetic excitations and electronic ground state of hydrogenated  CeNiSnH$ _x$. Here we report inelastic neutron scattering measurements on polycrystalline samples of CeNiSnD and CeNiSnD$_{1.8}$, as well as muon-spin relaxation ($\mu$SR) measurements of CeNiSnD. The $\mu$SR measurements confirm long-range magnetic order in CeNiSnD  where the whole sample volume orders magnetically below  $T_{\rm N}$. INS reveals well-defined CEF excitations in both CeNiSnD and CeNiSnD$_{1.8}$, demonstrating that hydrogenation leads to a localization of the Ce-4$f$ electrons in CeNiSn, and CEF level schemes are proposed.

\section{EXPERIMENTAL DETAILS}
 INS and $\mu$SR measurements were performed on powder samples of CeNiSnD$_x$ (x = 1, 1.8)  at the ISIS facility at the Rutherford Appleton Laboratory, UK. Note that deuterated, rather than hydrogenated, samples were utilized due to the much larger incoherent neutron scattering cross-section of hydrogen. INS were measured using the HET time-of-flight chopper spectrometer with incident neutron energies of 20 meV and 60 meV. On HET, neutrons are scattered from the sample into two forward detector banks at low scattering angles covering 2.4$^{\circ}$ $\rightarrow$ 6.9$^{\circ}$ and 9.0$^{\circ}$ $\rightarrow$ 29.0$^{\circ}$, and banks at 110.2 $^{\circ}$ $\rightarrow$ 119.2$^{\circ}$ and 125.1$^{\circ}$ $\rightarrow$ 138.4$^{\circ}$ at high scattering angles. In order to estimate the phonon contribution to the scattering of CeNiSnD$_{1.8}$,  isostructural  LaNiSnD$_{1.8}$ was also measured, while the phonon contribution for CeNiSnD was estimated from INS measurements of LaNiSn \cite{INSofdopedCeNiSn}. The magnetic contributions to the INS were fitted to a CEF model using Mantid \cite{mantidplots}.

\begin{figure}
	\includegraphics[angle=0,width=0.5\textwidth]{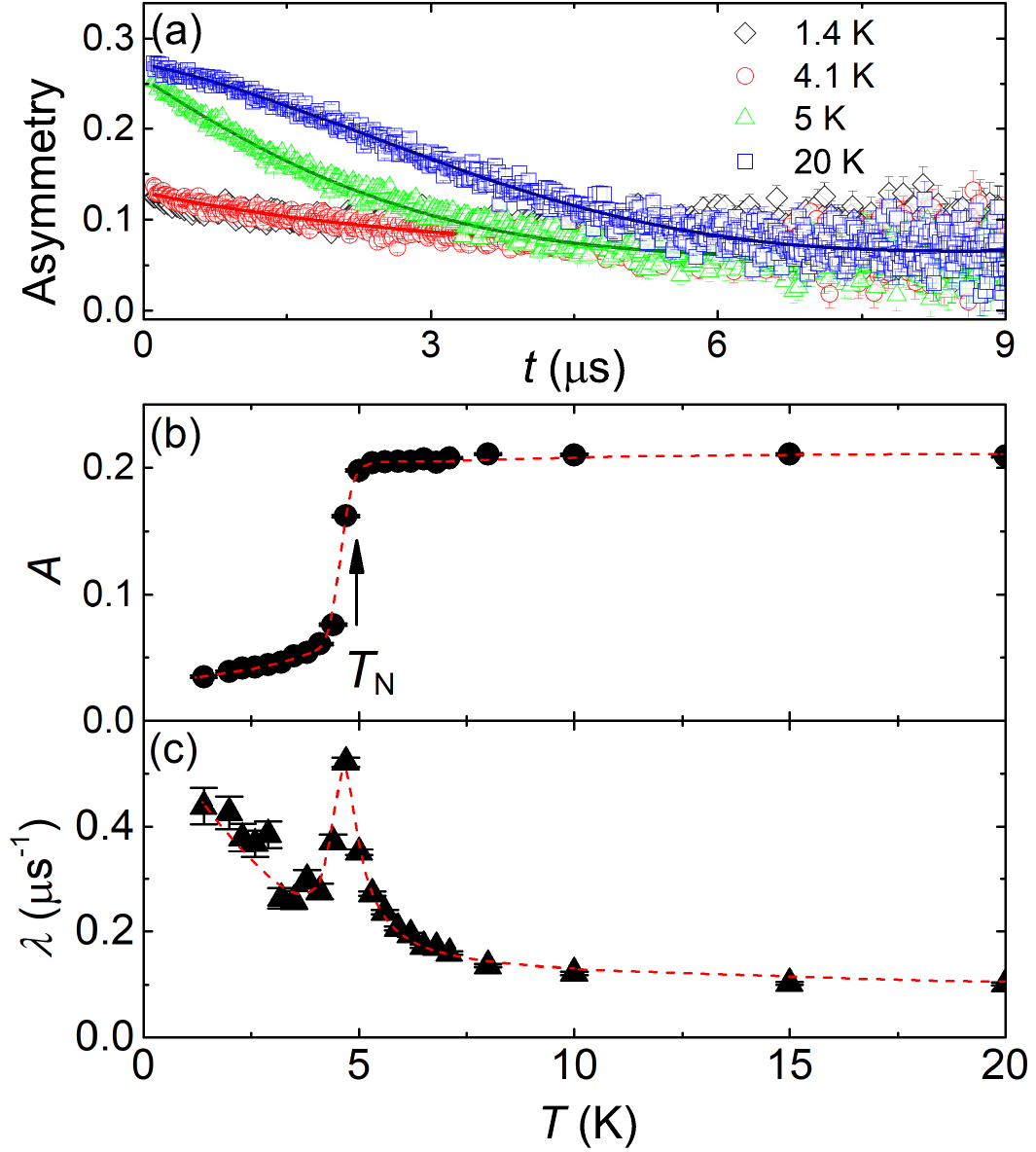}
	\caption{\label{Fig1}(a) Zero-field $\mu$SR spectra of CeNiSnD measured at four temperatures, both above and below the magnetic transition. The solid lines correspond to the fitting using Eq.~\ref{equa.1}, where (b) shows the temperature dependence of the  initial asymmetry $A$ of the component corresponding to the sample and (c) shows the Lorentzian relaxation rate $\lambda $. The red dashed lines in (b) and (c) are guides to the eye.}
	\vspace{-12pt}
\end{figure}

\section{RESULTS AND DISCUSSION}
\subsection{Zero-field $\mu$SR measurements}
Zero-field (ZF) muon spin relaxation measurements of CeNiSnD at selected temperatures are displayed in Fig.~\ref{Fig1}(a). At lower temperatures there is a significant drop in the asymmetry, consistent with the onset of long-range magnetic order. We do not observe signatures of coherent oscillations in the asymmetry  below $T_{\rm N}$, indicating that the local magnetic fields at the muon stopping site are too large for the corresponding oscillations to be resolved. Note that the width of the implanted muon pulse is around 80~ns at the ISIS Facility (compared to a time width of $<1$~ns for the continuous muon beam at the Paul Scherrer Institut), meaning that frequencies above around 10~MHz cannot be resolved. Since the spins of the implanted muons precess at the Larmor frequency associated with the perpendicular component of the local magnetic field, coherent oscillations associated with internal fields above around $\sim 700$~G will not be detected. On the other hand, coherent oscillations are observed in $\mu$SR measurements of both CeNiSnH and CeNiSnH$_{1.8}$ using a continuous muon source, which has much greater time resolution and allows measurements of GHz frequencies and fast relaxation rates ($>$100 $\mu$s$^{-1}$)\cite{CeNiSnHxpressure}.

The asymmetry were fitted to a damped Gaussian Kubo-Toyabe (KT) function
\begin{equation}
\label{equa.1}
G_{ZF}(t)=A\bigg[ \frac{1}{3}+\frac{2}{3}(1-\delta^2 t^2){\rm exp}\Bigg(-\frac{\delta^2 t^2}{2}\Bigg)\bigg]{\rm exp}(-\lambda t)+A_{bg},
\end{equation}
where $A$ is the initial asymmetry, $A_{bg}$ corresponds to the time-independent background term from muons stopping in the silver sample holder, while $\delta$ and $\lambda$ are the Gaussian and exponential relaxation rates, respectively. The value of $\delta$ was fixed from fitting the data at 20 K, while the temperature dependence of the fitted $A$ is displayed in Fig~\ref{Fig1}(b). There is a sharp drop in $A$ below 5~K, which decreases to a value of around one third of that at high temperatures. Such a loss of asymmetry in measurements of polycrystalline samples strongly suggests that the whole sample magnetically orders below the magnetic transition.  There is also  a sharp peak in $\lambda$ at 4.7~K  [Fig.~\ref{Fig1}(c)], which is consistent with the critical slowing down of spin fluctuations upon approaching the  transition.

\subsection{Inelastic neutron scattering}
\subsubsection{CeNiSnD$_{1.8}$}
\begin{figure}
	\includegraphics[angle=0,width=0.5\textwidth]{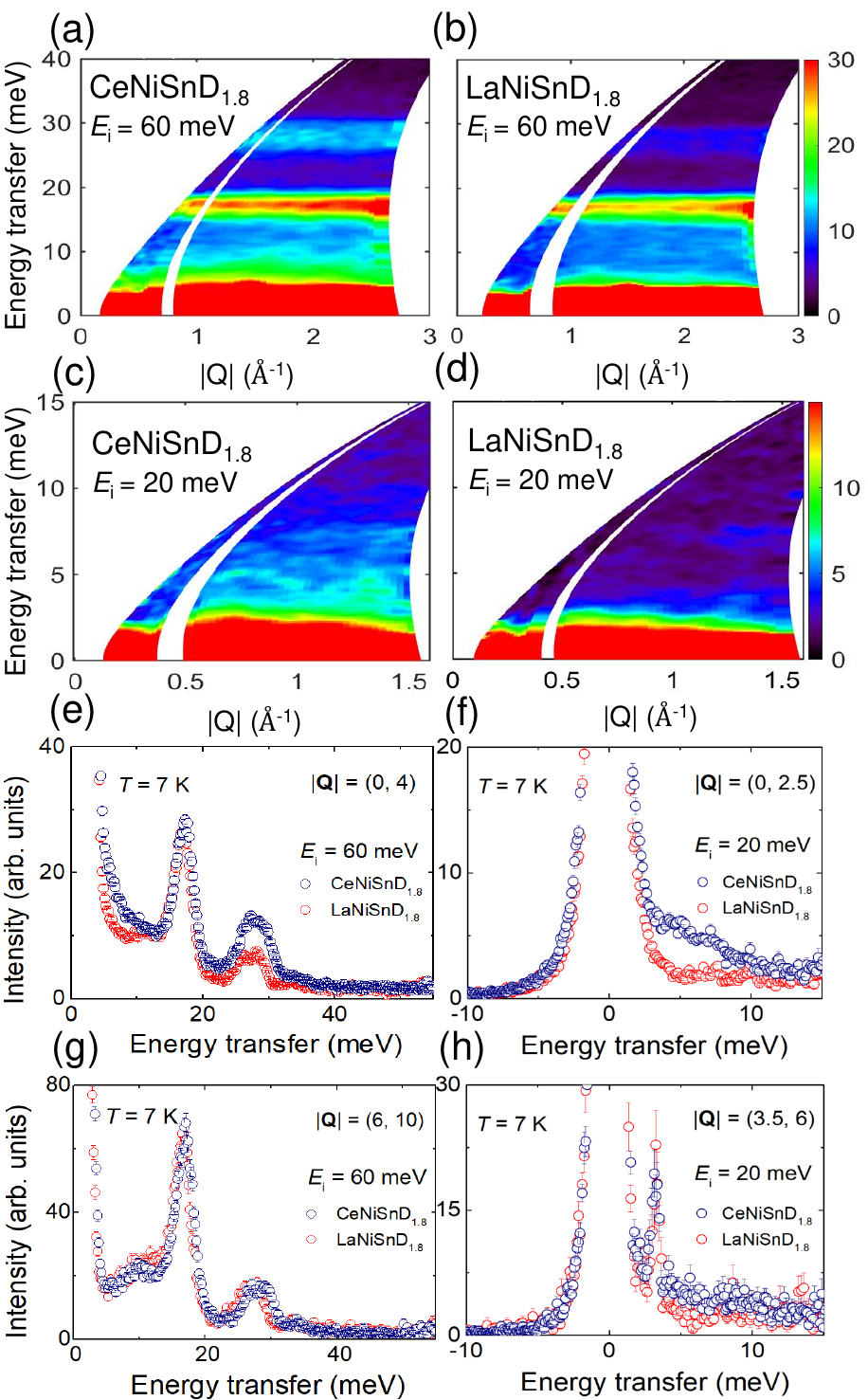}
	\caption{\label{Fig2} Color plots of the inelastic neutron scattering response in the low angle banks as a function of energy transfer $E$ and momentum transfer $\vert\mathbf{Q}\vert$ for (a) CeNiSnD$_{1.8}$ at 7 K  ($E_{\rm i}$ = 60 meV), (b) LaNiSnD$_{1.8}$ at 7 K  ($E_{\rm i}$ = 60 meV), (c) CeNiSnD$_{1.8}$ at 7 K ($E_{\rm i}$ = 20 meV), and (d) LaNiSnD$_{1.8}$ at 7 K ($E_{\rm i}$ = 20 meV). Cuts of the scattering intensity of CeNiSnD$_{1.8}$ and LaNiSnD$_{1.8}$ at 7~K versus $E$ obtained from integrating across low-$\vert\mathbf{Q}\vert$ in the range (e) 0-4 $\rm{\AA^{-1}}$ with $E_{\rm i}$ = 60 meV, and (f) 0-2.5 $\rm{\AA^{-1}}$ with $E_{\rm i}$ = 20 meV. Corresponding high-$\vert\mathbf{Q}\vert$ cuts are also displayed in the range (g) 6-10 $\rm{\AA^{-1}}$ with $E_{\rm i}$ = 60 meV, and (h) 3.5-6 $\rm{\AA^{-1}}$ with $E_{\rm i}$ = 20 meV.  Note that for $E_{\rm i}$ = 60 meV, the low angle banks of the HET spectrometer reach a maximum momentum transfer of $3~\rm{\AA^{-1}}$ for energy transfers below 40~meV, but to around $4~\rm{\AA^{-1}}$ at higher energies.} 
	\vspace{-12pt}
\end{figure}

The inelastic neutron scattering responses of CeNiSnD$_{1.8}$ and the non-magnetic analog LaNiSnD$_{1.8}$ measured with an incident energy $E_{\rm i}=60$~meV at 7~K are shown as color-coded intensity plots in Figs. \ref{Fig2}(a) and \ref{Fig2}(b)  respectively, while the responses  with $E_{\rm i}=20$~meV at $T$ = 7 K are shown in Figs. \ref{Fig2}(c) and \ref{Fig2}(d). A clear dispersionless excitation is observed at around 27 meV, while there is only very weak intensity at this energy transfer for LaNiSnD$_{1.8}$, and therefore this can be ascribed to be a crystalline-electric field excitation of the Ce$^{3+}$ ions. There is also additional intensity at low momentum transfers $\vert$$\mathbf{Q}$$\vert$ at low energy close to the elastic line in CeNiSnD$_{1.8}$, which corresponds to magnetic scattering. This can be clearly seen from a comparison of the cuts integrated over low $\vert$$\mathbf{Q}$$\vert$, corresponding to momentum ranges of 0-4 $\rm{\AA^{-1}}$ for $E_{\rm i}=60$~meV  and 0-2.5 $\rm{\AA^{-1}}$ for $E_{\rm i}=20$~meV, displayed in Figs. \ref{Fig2}(e)  and \ref{Fig2}(f), respectively,  where the intensity of the CeNiSnD$_{1.8}$ data exceeds that of  LaNiSnD$_{1.8}$ both in the vicinity of the 27 meV excitation, and below 10~meV.  On the other hand,  Figs. \ref{Fig2}(g)  and \ref{Fig2}(h) show cuts obtained from integrating over high $\vert$$\mathbf{Q}$$\vert$, for $E_{\rm i}=60$~meV (6-10 $\rm{\AA^{-1}}$) and $E_{\rm i}=20$~meV (3.5-6 $\rm{\AA^{-1}}$), respectively. It can be seen that these cuts are very similar between CeNiSnD$_{1.8}$ and LaNiSnD$_{1.8}$, suggesting very similar phonon spectra between the two systems.

\begin{figure}
	\includegraphics[angle=0,width=0.42\textwidth]{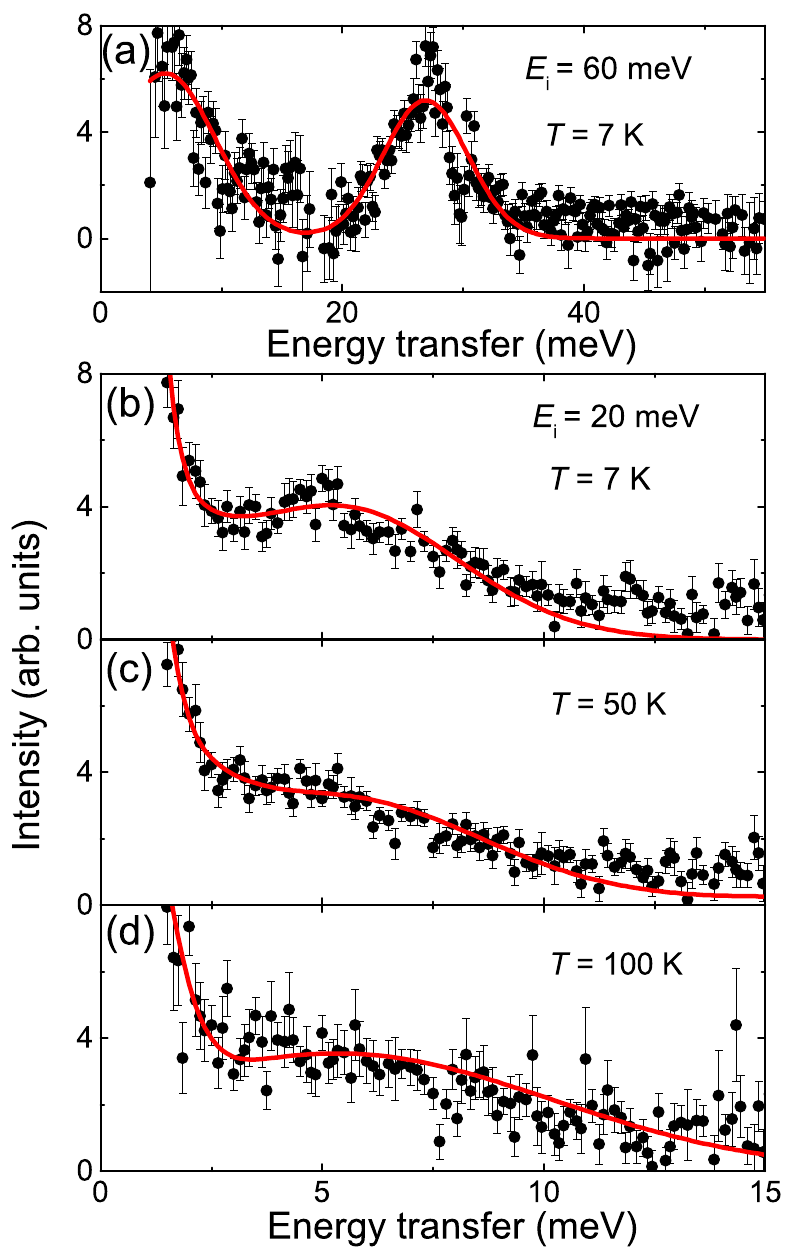}
	\caption{\label{Fig3}Estimated magnetic contribution to the inelastic neutron scattering intensity of CeNiSnD$_{1.8}$ for (a) $ E_{\rm i} $ = 60 meV at 7 K, (b) $ E_{\rm i} $ = 20 meV at 7 K, (c) $ E_{\rm i} $ = 20 meV at 50 K, and (d) $ E_{\rm i} $ = 20 meV at 100 K. The solids lines show a fit to the CEF model described in the text.}
	\vspace{-12pt}
\end{figure}

Figure~\ref{Fig3} displays cuts of the estimated intensity of the magnetic scattering ($S_{\rm {mag}}^{\rm {Ce}}$) versus energy transfer. The magnetic contribution for $E_{\rm i}$ = 60 meV are obtained by subtracting an estimate of the phonon contribution  of LaNiSnD$_{1.8}$ using
\begin{equation}
S_{\rm {mag}}^{\rm {Ce}}(\omega) = S_{\rm {low}}^{\rm {Ce}}(\omega) - S_{\rm {high}}^{\rm {Ce}}(\omega)  \bigg[S_{\rm {low}}^{\rm {La}}(\omega)/S_{\rm {high}}^{\rm {La}}(\omega)\bigg] 
\end{equation} 
\noindent where the $S_{\rm {low}}$ and $S_{\rm {high}}$ correspond to cuts integrated over low and high  $\vert$$\mathbf{Q}$$\vert$, respectively. For  $E_{\rm i}$ = 20 meV, the high $\vert$$\mathbf{Q}$$\vert$ data of LaNiSnD$_{1.8}$ has relatively poor statistics, and therefore the magnetic contribution was estimated via directly subtracting the data of LaNiSnD$_{1.8}$ using
\begin{equation}
\label{Eq2}
S_{\rm {mag}}^{\rm {Ce}}(\omega) = S_{\rm {low}}^{\rm {Ce}}(\omega) - \alpha S_{\rm {low}}^{\rm {La}}(\omega),
\end{equation}

\noindent where $\alpha$ = 0.835 is the ratio of the total neutron scattering cross sections per formula unit of CeNiSnD$_{1.8}$ and LaNiSnD$_{1.8}$. These data support there being CEF excitations at around 5 meV and 27 meV in CeNiSnD$_{1.8}$. In the hexagonal crystal structure of  CeNiSnD$_{1.8}$ with space group $P6_3/mmc$ (No. 194), the Ce ions have trigonal point symmetry (D$_{3d}$) \cite{CeNiSnH1.8,CEFCeNiSnH1.8}, and therefore the corresponding CEF Hamiltonian ($H_{\rm {CEF}}$) has the  form:
\begin{equation}
H_{\rm {CEF}} = B_2^0O_2^0 + B_4^0O_4^0+B_4^3O_4^3 
\end{equation}
where $B_n^m$ are Stevens parameters and $O_n^m$ are the Stevens operator equivalents \cite{KWHStevens1952}.  As shown by the solid lines in Fig.~\ref{Fig3}, the data can be well fitted by this CEF model. Here, the 7~K data  with both $E_{\rm i}$ = 60 meV and 20 meV  were simultaneously fitted to obtain the CEF parameters. These parameters were then fixed in order to fit the $E_{\rm i}$ = 20 meV data at 50 and 100~K, which are also well described by this model. The obtained parameters are $B_2^0=0.802$~meV, $B_4^0=0.022$~meV, and $B_4^3=-1.237$~meV. This gives rise to three doublets, where the energy differences between the ground state and the two excited levels are 5.34~meV and 26.91~meV and the corresponding wave functions of the three doublets are shown in Table \ref{wavefunction}. Here the wave function of the ground-state doublet ($\vert m_J\rangle$) is $\vert\psi^{\pm}\rangle$ = -0.863$\vert\pm\frac{1}{2}\rangle$ $\pm$ 0.505$\vert\mp\frac{5}{2}\rangle$. 

These results contrast with the CEF scheme with a  $\vert\frac{3}{2}\rangle$ ground state, proposed based solely on magnetization measurements of CeNiSnH$_{1.8}$ polycrystalline samples \cite{CeNiSnH1.8}. For such a CEF level scheme, only the out-of-plane magnetic moment is non-zero, and the predicted splitting of the excited CEF states is too small to account for our INS results. On the other hand, for our CEF scheme the predicted ground-state magnetic moments along the $c$~axis and in the $ab$~plane are $\mu_{\rm z}$ = 0.23$\mu_{\rm B}$/Ce and $\mu_{\rm x}$ = 0.96$\mu_{\rm B}$/Ce, respectively, and an expected magnetization of 0.73$\mu_{\rm B}$/Ce is calculated for a polycrystalline sample in a 5 T applied field, which is slightly larger than the observed value of 0.5-0.6$\mu_{\rm B}$/Ce \cite{CeNiSnH1.8,Fernandez2009}. Furthermore, our deduced CEF ground state predicts a magnetic moment within the $ab$ plane, in contrast to the uniaxial anisotropy of the CEF scheme in Ref.~\cite{CeNiSnH1.8}, but experimental determinations of the ordered moment direction, either from neutron diffraction or characterization of single crystals, are still lacking.

\begin{figure}
	\includegraphics[angle=0,width=0.5\textwidth]{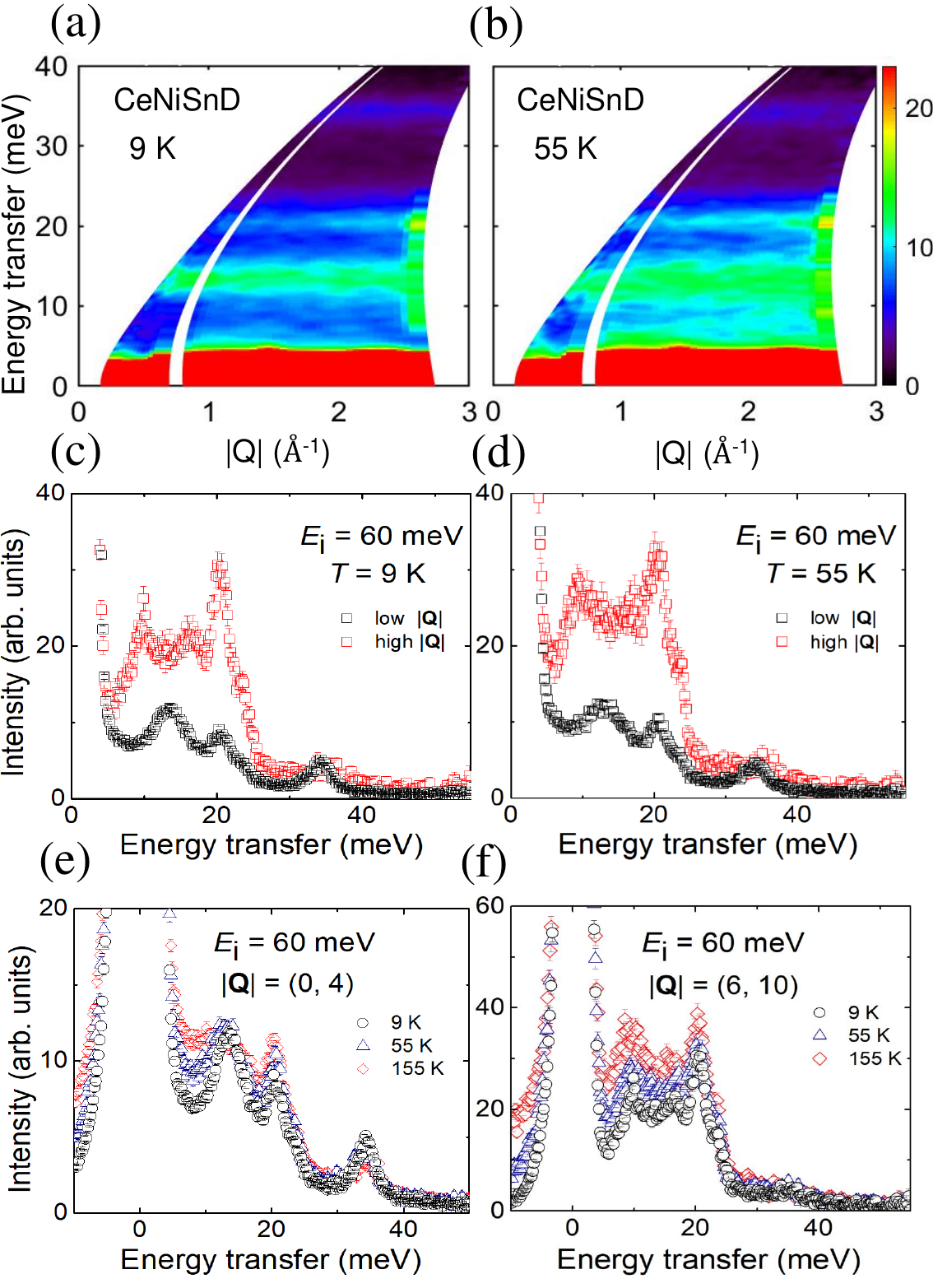}
	\caption{\label{Fig4}Color-coded plot of the inelastic neutron scattering intensity in the low angle banks as a function of energy and momentum transfer for CeNiSnD at (a) 9 K and (b) 55 K, with $E_{\rm i}$ = 60 meV. The scattering intensity scale is in arbitrary units. Cuts of the neutron scattering intensity as a function of energy transfer obtained from integrating across a low-$\vert$$\mathbf{Q}$$\vert$  (0-4 $\rm{\AA^{-1}}$) and a high-$\vert$$\mathbf{Q}$$\vert$ (6-10 $\rm{\AA^{-1}}$) range for $E_{\rm i}$ = 60 meV at (c) 9 K and (d) 55 K. Cuts at various temperatures for $E_{\rm i}$ = 60 meV are displayed obtained from integrating across  (e) low-$\vert$$\mathbf{Q}$$\vert$, and (f) high-$\vert$$\mathbf{Q}$$\vert$ ranges.  }
	\vspace{-12pt}
\end{figure}

\subsubsection{CeNiSnD}
Inelastic neutron scattering measurements were also performed on CeNiSnD at several temperatures  above $T_{\rm N}$ with  $E_{\rm i}=60$~meV, in order to investigate the CEF excitations. Two-dimensional color plots of the scattering intensity versus $E$ and $\vert$$\mathbf{Q}$$\vert$  at 9 and 55 K are displayed in Figs. \ref{Fig4}(a) and \ref{Fig4}(b), respectively. 
Corresponding cuts of the intensity of the data at the two temperatures integrated over low  $\vert$$\mathbf{Q}$$\vert$ (0-4 $\rm{\AA^{-1}}$) and high $\vert$$\mathbf{Q}$$\vert$ (6-10 $\rm{\AA^{-1}}$) are displayed in Figs. \ref{Fig4}(c) and \ref{Fig4}(d). It can be seen in these cuts that at around 34~meV there is a pronounced peak in the low  $\vert$$\mathbf{Q}$$\vert$ data of CeNiSnD that is absent at high   $\vert$$\mathbf{Q}$$\vert$. Meanwhile, while the high  $\vert$$\mathbf{Q}$$\vert$ data below 30~meV are larger than at low  $\vert$$\mathbf{Q}$$\vert$ due to  phonon contributions, there is also a peak in the  low  $\vert$$\mathbf{Q}$$\vert$ data at around 13~meV that is absent at high   $\vert$$\mathbf{Q}$$\vert$. The intensity of these two magnetic excitations decrease with increasing temperature, as shown in Fig. \ref{Fig4}(e), consistent with these corresponding to CEF excitations from the ground state doublet to excited levels, while the peaks in the high   $\vert$$\mathbf{Q}$$\vert$ data increase at higher temperatures [Fig. \ref{Fig4}(f)]. Moreover, it can be seen that with increasing temperature the intensity of the magnetic scattering at around 21~meV also increases, suggesting that this corresponds to a transition between the first and second excited CEF levels, once there is sufficient thermal energy to populate the first excited state. Note that a peak close to this energy is still present at 9~K, which has a larger intensity at high   $\vert$$\mathbf{Q}$$\vert$, suggesting that there is also a phonon excitation at similar energy transfers.

\begin{figure}
	\includegraphics[angle=0,width=0.44\textwidth]{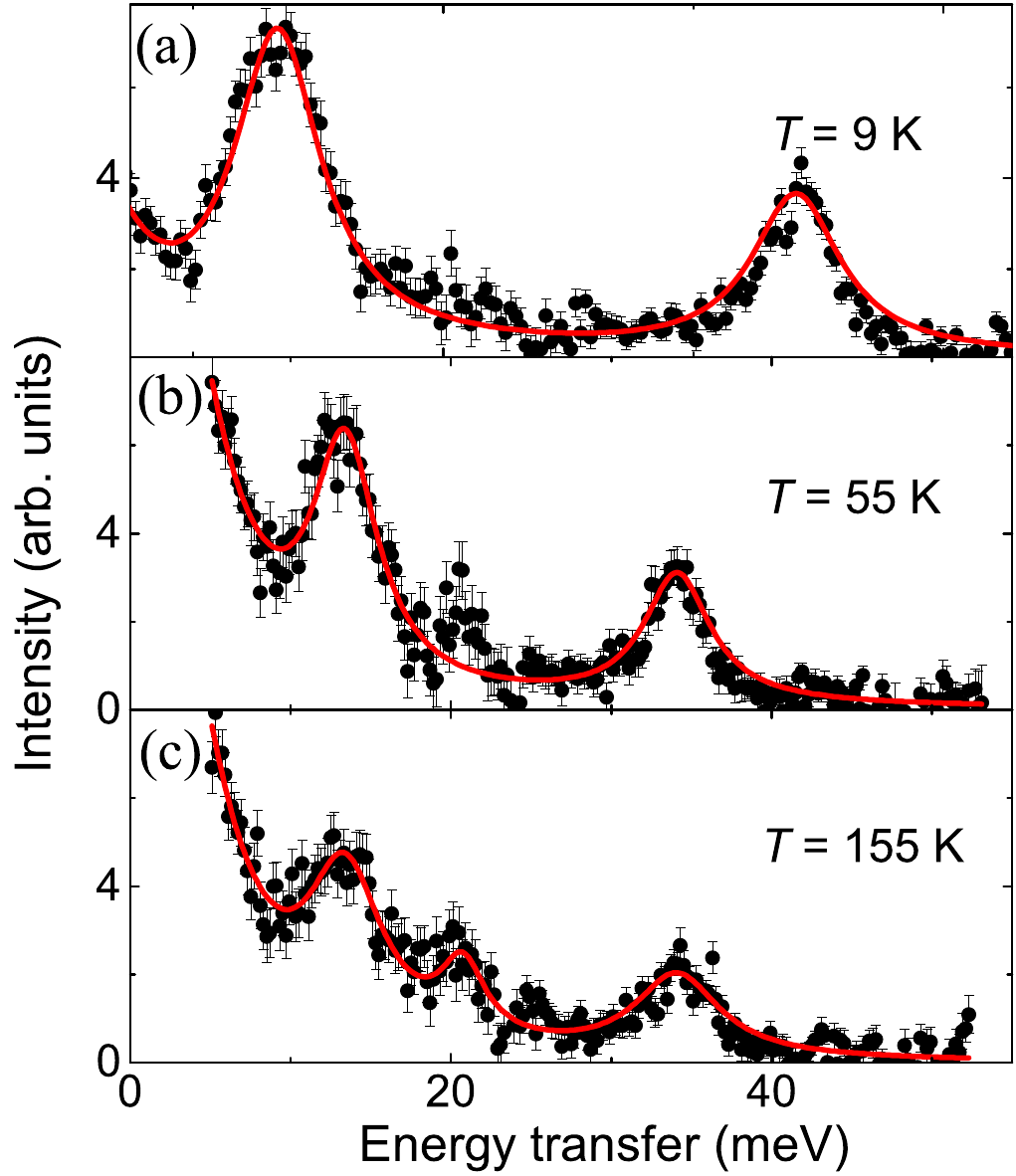}	
	\caption{\label{Fig5}Magnetic contribution to the inelastic neutron scattering intensity of CeNiSnD with  $E_{\rm i}$ = 60 meV at (a) 9~K, (b) 55~K, and (c) 155~K. The solids lines show a fit to the CEF model described in the text. Note that there is some additional intensity at around 20 meV at 55 K, which likely corresponds to phonon scattering.} 
	\vspace{-12pt}
\end{figure}

CeNiSnD crystallizes in an orthorhombic structure, which has the same crystal  structure as CeNiSn besides the addition of D$^+$ ions. Therefore in order to estimate the magnetic scattering from CeNiSnD, the data of non-magnetic LaNiSn \cite{INSofdopedCeNiSn} were directly subtracted using Eq.~\ref{Eq2}, and are displayed in Fig. \ref{Fig5} at three temperatures. It can be seen that at 9~K there are two pronounced peaks corresponding to the aforementioned CEF excitations at around 13 and 34 meV. Note also that some additional intensity can still be detected at around 20~meV, which is likely a result of an incomplete subtraction of the phonon contribution. Since CeNiSnD contains light D atoms, which are absent in non-magnetic LaNiSn, it might be expected that the phonon spectra of these compounds should be similar at low energies, but there could be some difference at higher energies.

For  Ce$^{3+}$ in an orthorhombic CEF with local point group symmetry $C_1$, the $J = 5/2$ ground state multiplet is also expected to split into three Kramers doublets in the paramagnetic state. The corresponding Hamiltonian  is given by 

\begin{equation}
\label{equa.5}
\begin{split}
 H_{\rm {CEF}} = &B_2^0O_2^0+B_2^{\pm 1}O_2^{\pm 1}+B_2^{\pm 2}O_2^{\pm 2}+B_4^0O_4^0+B_4^{\pm 1}O_4^{\pm 1} \\
&+B_4^{\pm 2}O_4^{\pm 2}+B_4^{\pm 3}O_4^{\pm 3}+B_4^{\pm 4}O_4^{\pm 4} 
\end{split}
\end{equation}

\begin{table}
	\centering
	\caption{Crystal electric field parameters $B_n^{\pm m}$ obtained from the analysis of the inelastic neutron scattering data of CeNiSnD.}
	\label{CEFparameters}
	\begin{tabular}{cc}
		\toprule
		\multicolumn{1}{m{3cm}}{\centering $B_n^{\pm m}$} & \multicolumn{1}{m{3cm}}{\centering Value (meV)}\\
		\midrule
		$B_2^0$ & 0.102\\
		$B_2^1$ & 0.428\\
		$B_2^2$ & 0.895\\
		$B_2^{-1}$ & -1.152\\
		$B_2^{-2}$ & -2.029\\
		$B_4^0$ & 0.009\\
		$B_4^1$ & 0.020\\
		$B_4^2$ & 0.035\\
		$B_4^3$ & -0.098\\
		$B_4^4$ & -0.125\\
		$B_4^{-1}$ & -0.019\\
		$B_4^{-2}$ & -0.319\\
		$B_4^{-3}$ & -0.035\\
		$B_4^{-4}$ & -0.044\\
		\bottomrule
	\end{tabular}
\end{table}

Due to low point group symmetry of the Ce ions, there are a large number of  CEF parameters $B_n^{\pm m}$ with 14 independent CEF parameters, which could not be uniquely determined from the observed spectra with two CEF excitations. Therefore we fitted the data with an initial set of parameters corresponding to a point charge model with CEF excitations very close to those observed \cite{PCMPr2Pd3Ge5}. Using these initial parameters, the  magnetic scattering  at 9 K, 55 K and 155 K were simulatenously fitted, and the resulting parameters are shown in Table \ref{CEFparameters}. It can be seen that this model can well fit the data, where there are excitations from the ground state to excited state doublets at 13.4 meV and 34.1 meV, while at high temperatures there is an addition excitation corresponding to the transition between the first and second excited states. The corresponding wave functions of the three doublets are shown in Table \ref{wavefunction}. Based on this CEF scheme, the simulated magnetization predicts the $c$~axis being the easy axis with a saturated moment of 1.26$\mu_{\rm B}$/Ce which is consistent with the observed $c$~axis ordered moment of 1.37 $\mu_{\rm B}$/Ce deduced from neutron diffraction \cite{PNDofCeNiSnD}.  However, given the large number of CEF parameters, this can only be regarded as a possible CEF scheme, and further measurements would be necessary to further constrain the CEF parameters, such as  polarized neutron scattering on single crystal samples, as performed for CePtSn \cite{PhysRevB.69.220412}.

\begin{table*}[htp]
	\centering
	\caption{Energy levels and associated wave functions obtained from the analysis of the neutron scattering data of CeNiSnD$_{1.8}$ and CeNiSnD using a CEF model.}
	\label{wavefunction}
	\renewcommand\arraystretch{1.25}
	\begin{tabular}{ccc}
		\toprule
		\multicolumn{2}{c}{CeNiSnD$_{1.8}$} \\ \cline{1-2}
		\multicolumn{1}{m{3cm}}{\centering Energy levels (meV)} & 
		\multicolumn{1}{m{14cm}}{\centering Wave functions}\\
		\midrule	
		0 &-0.863$\vert\pm\frac{1}{2}\rangle$ $\pm$ 0.505$\vert\mp\frac{5}{2}\rangle$\\
		5.338 & 
		$\vert\pm\frac{3}{2}\rangle$\\
		26.909 & 
		0.505$\vert\mp\frac{1}{2}\rangle$ $\mp$ 0.863$\vert\pm\frac{5}{2}\rangle$\\
		\midrule
		\\
		\\
		\toprule
		\multicolumn{2}{c}{CeNiSnD} \\ \cline{1-2}
		\multicolumn{1}{m{3cm}}{\centering Energy levels (meV)} & 
		\multicolumn{1}{m{14cm}}{\centering Wave functions}\\
		\midrule
		0 & 0.681$\vert-\frac{5}{2}\rangle$+(0.008+i0.144)$\vert-\frac{3}{2}\rangle$-(0.210+i0.664)$\vert-\frac{1}{2}\rangle$-(0.035+i0.006)$\vert\frac{1}{2}\rangle$+(0.121+i0.117)$\vert\frac{3}{2}\rangle$\\
		0 & (-0.158+i0.057)$\vert-\frac{3}{2}\rangle$-(0.034+i0.008)$\vert-\frac{1}{2}\rangle$+(0.463-i0.521)$\vert\frac{1}{2}\rangle$+(0.066-i0.128)$\vert\frac{3}{2}\rangle$-(0.622+i0.278)$\vert\frac{5}{2}\rangle$\\
		13.385 & (0.869+i0.242)$\vert-\frac{3}{2}\rangle$+(0.122+i0.017)$\vert-\frac{1}{2}\rangle$+(0.110-i0.213)$\vert\frac{1}{2}\rangle$+(-0.202+i0.174)$\vert\frac{3}{2}\rangle$+(0.084-i0.193)$\vert\frac{5}{2}\rangle$\\
		13.385 & -0.211$\vert-\frac{5}{2}\rangle$+(-0.240+i0.116)$\vert-\frac{3}{2}\rangle$-(0.239-i0.016)$\vert-\frac{1}{2}\rangle$+(0.029-i0.109)$\vert\frac{1}{2}\rangle$+(-0.125+i0.894)$\vert\frac{3}{2}\rangle$\\
		34.098 & (-0.168+i0.141)$\vert-\frac{3}{2}\rangle$+(-0.030+i0.040)$\vert-\frac{1}{2}\rangle$-(0.636+i0.190)$\vert\frac{1}{2}\rangle$-(0.097+i0.089)$\vert\frac{3}{2}\rangle$+(0.063-i0.698)$\vert\frac{5}{2}\rangle$\\
		34.098 & 0.701$\vert-\frac{5}{2}\rangle$-(0.080+i0.105)$\vert-\frac{3}{2}\rangle$+(0.132+i0.650)$\vert-\frac{1}{2}\rangle$+(0.042-i0.027)$\vert\frac{1}{2}\rangle$-(0.155-i0.155)$\vert\frac{3}{2}\rangle$\\
		\midrule
		
	\end{tabular}
\end{table*}

\section{SUMMARY}
 We probed powder samples of CeNiSnD and CeNiSnD$_{1.8}$ using zero field muon spin relaxation and inelastic neutron scattering measurements. $\mu$SR measurements of of CeNiSnD confirm the presence of long-range magnetic order, whereby the whole sample undergoes a magnetic transition below $T_{\rm N}$ = 4.7 K. Moreover the inelastic neutron scattering measurements reveal the presence of two well-defined CEF excitations in both compounds, at around 13 meV and 34 meV in CeNiSnD, and 5 meV and 27 meV for CeNiSnD$_{1.8}$. This is in contrast to CeNiSn, where evidence is found for a weak broad possible CEF excitation near 40 meV \cite{CEFCeNiSn,possibleCEFCeNiSn}. These results therefore suggest that the insertion of deuteurium to CeNiSn decreases the hybridization between the 4$f$ electrons and the conduction electrons, leading to the localization of the 4$f$ electrons, together with the occurrence of long range magnetic order. This is consistent with the lack of magnetic order in CeNiSn being a consequence of the strong Kondo hybridization \cite{CeNiPdSn1991,CEFCePtSnNidoped}, rather than there being a disordered arrangement of local moments. The 40~meV excitation in CeNiSn is slightly higher in energy than the second excited level of isostructural CeNiSnD suggesting that the CEF potential is either modified by the presence of deuteurium ions, or the different nature of the conduction electrons. A CEF model for the trigonal local symmetry of the Ce ion can well account for the magnetic scattering of CeNiSnD$_{1.8}$, for which the predicted moment for the resulting ground state is 0.96$\mu_{\rm B}$/Ce within the $ab$-plane, and the observed magnetization for a polycrystalline sample is similar to that expected for the CEF model  \cite{CeNiSnH1.8,Fernandez2009}. However, determining whether there is a reduced ordered moment arising from Kondo coupling between the 4$f$ and conduction electrons requires further characterization of the magnetic ground state of CeNiSnD$_{1.8}$.  Meanwhile we propose a tentative CEF model for CeNiSnD, for which the saturated moment  along the easy $c$ axis of 1.26$\mu_{\rm B}$/Ce is also similar to that deduced from neutron diffraction \cite{PNDofCeNiSnD}, but the low symmetry of the Ce site means that additional measurements would be necessary to constrain the CEF parameters.

\begin{acknowledgments}
This work was supported by the National Key R$\&$D Program of China (Grants No. 2022YFA1402200, and No. 2023YFA1406303), the Key R$\&$D Program of Zhejiang Province, China (Grant No. 2021C01002),
the National Natural Science Foundation of China (Grants No. 12222410, No. 11974306, No. 12034017, and No. 12174332), and the Zhejiang Provincial Natural Science Foundation of
China (Grant No. LR22A040002). D.T.A. would like to
thank the Royal Society of London for International Exchange
funding between the UK and Japan, Newton Advanced Fellowship funding between UK and China, and EPSRC UK for the funding (Grant No. EP/W00562X/1).
\end{acknowledgments}

\end{document}